\documentclass[11pt,a4paper]{emulateapj}

\usepackage{epsfig}
\usepackage{amsmath}
\usepackage{natbib}
\usepackage{multirow}

\def\bb{$B_{\rm 435}$} 
\def\vv{$V_{\rm 606}$} 
\def\ii{$i_{\rm 775}$} 
\def\zz{$z_{\rm 850}$} 

\shorttitle{Surface Brightness Dimming}
\shortauthors{Calvi V. et al.}

\begin{document}

\title{The Effect of Surface Brightness Dimming in the Selection of
  High-$z$ Galaxies} \author{V. Calvi\altaffilmark{1,2},
  M. Stiavelli\altaffilmark{1}, L. Bradley\altaffilmark{1},
  A. Pizzella\altaffilmark{2,3}, S. Kim\altaffilmark{4}} \date{\today}

\altaffiltext{1}{Space Telescope Science Institute, 3700 San Martin
  Drive, Baltimore, MD 21218, USA. Email: calvi@stsci.edu}
\altaffiltext{2}{Dipartimento di Fisica e Astronomia ``G. Galilei'',
  vicolo dell'Osservatorio 3, 35122, Padova (PD), Italy}
\altaffiltext{3}{INAF-Osservatorio Astronomico, vicolo
  dell'Osservatorio 5, 35122, Padova (PD), Italy} \altaffiltext{4}{The
  Johns Hopkins University, 3400 North Charles Street, Baltimore, MD
  21218, USA}

\begin{abstract}
Cosmological surface brightness dimming of the form $(1+z)^{-4}$
affects all sources. The strong dependence of surface brightness
dimming on redshift $z$ suggests the presence of a selection bias when
searching for high-redshift galaxies, i.e. we tend to detect only
those galaxies with a high surface brightness (SB).  However,
unresolved knots of emission are not affected by SB dimming,  thus
providing a way to test the clumpiness of high-$z$ galaxies. Our
strategy relies on the comparison of the total flux detected for the
same source in surveys characterized by different depth.  For all
galaxies, deeper images permit the better investigation of low-SB
features. Cosmological SB dimming makes these low-SB features hard to
detect when going to higher and higher redshifts.   We used  the
GOODS and HUDF Hubble Space Telescope legacy datasets to study the
effect of SB dimming on low-SB features of high-redshift galaxies and
compare it to the prediction for smooth sources.  We selected a sample
of Lyman-break galaxies at $z\sim4$ (i.e. \bb-band dropouts) detected
in all of the datasets and found no significant trend when comparing
the total magnitudes measured from images with different depth. 
  Through Monte Carlo simulations we derived the expected trend for
  galaxies with different SB profiles. The comparison to the data
hints at a compact distribution for most of the rest-frame ultraviolet
light emitted from high-$z$ galaxies.

\end{abstract}

\keywords{galaxies: evolution --- galaxies: high-redshift ---
  galaxies: photometry}

\section{Introduction}
\label{sec:introduction}

The surface brightness (SB) dimming derived by
\cite{tolman1930,tolman1934} affects all cosmological sources.  In
particular, since for every extended source the surface brightness $I$
is no longer redshift-independent as in an Euclidean Universe, there
is an additional term that should be taken into account. The
contributions of time dilation, redshift, and curvature make high-$z$
galaxies progressively difficult to detect according to
\begin{equation}
I_0 =\frac{I_e}{(1+z)^4}
\end{equation}
 where $I_0$ and $I_e$ are the observed and intrinsic surface
  brightnesses of the object, respectively.
In particular, this effect is relevant for sources at $z \ge 4$ that
dim by a factor greater than 600.  The observational effect of
cosmological dimming is a bias in the detection of very high-$z$
galaxies, i.e. we tend to detect only those galaxies with a high
surface brightness.  The dimming effect affects
our understanding of the reionization phase as well, since the role of
galaxies in this process and, in particular, the amount of ionizing
radiation produced depends directly on the number of high-$z$
candidates detected in deep surveys.

The first studies on cosmological dimming and the consequent
observational bias towards more compact objects at higher redshifts
were conducted by \cite{phillips1990}, but only up to redshift about
0.3. Later, \cite{pascarelle1998} and \cite{lanzetta2002} stated that a
correction is needed in order to sample high and low-$z$ galaxies at
the same SB threshold and that only the bright regions of galaxies are
observable at high-$z$, since the faint ones cannot be detected
against the background.  A comprehensive discussion on Tolman dimming
can be found in \cite{disney2012} who studied the surface brightness
selection in the framework of the search for progenitors of low-$z$
galaxies.  On the basis of Tolman dimming effect, they claim that,
looking back in time, there should be different kind of objects at
different epochs, in particular more compact objects at higher
redshifts.  Moreover, they claim the high-$z$ galaxy population,
mostly hidden because of dimming effects, could be the source of the
ionizing flux required to reionize the Universe . We will not discuss
this aspect here. Our comprehensive study of the role of faint
undetected galaxies in cosmic reionization is presented in
\cite{calvi2013}.

In the last 20 years, the Hubble Space Telescope (HST) has enabled the
astronomical community to study the morphology of Lyman Break Galaxies
(LBGs) at different redshifts, from $z\sim 1.5 $ up to $z\sim 7-8$,
and all those objects were found to be compact
\citep{giavalisco2004,lotz2006,oesch2010,williams2014}.
\cite{bouwens2004size} found that the principal effect
of depth in galaxy surveys is to add galaxies at fainter magnitudes,
with an approximate $(1+z)^{-1}$ trend on the sizes. In
particular, since galaxies at high-$z$ are mostly compact objects
\citep{bouwens2004size,oesch2010} the effect of cosmological dimming
is generally not expected to substantially affect their measured
fluxes when comparing surveys with different magnitude limit.

The goal of this work is to explore the effects of cosmological
surface brightness dimming in the selection of galaxies at
high-redshift. Our empirical strategy relies on the comparison of the
total flux detected for the same source in surveys characterized by
different depth and makes use of Monte Carlo simulations to derive the
expected trend when assuming different SB profiles.  To this end, we
considered datasets taken with HST that are characterized by different
depths. By using independent datasets, our results should be more
robust against statistical errors or systematics.  It should be noted
that we did not study the possible effect of photometric bias on the
color selection of high-$z$ candidates.

This paper is organized as follows: in Section 1 we will briefly
describe the datasets we used, in Section 2 we will focus  on
the construction of the rescaled rms maps that are required in the
galaxy selection process. Section 3 and 4 are devoted to the sample
selection procedure and the analysis that led to our results. In
Section 5 we will describe the Monte Carlo simulations we used to
constrain the expected trend from theory and, finally, in Section 6 we
will summarize our results and discuss our findings.

Throughout this paper we will use the  AB magnitude system
\citep{oke1983}.

\section{Datasets}
\label{Sec:dataset}

With the aim of a quantitative study on the surface brightness of
high-redshift galaxies selected on the basis of the Lyman break
technique \citep{steidel1999,giavalisco2002}, we compared results
obtained from different ultra deep images taken with the Advanced
Camera for Survey (ACS) installed in HST.  

In the following analysis, we will consider the Hubble Ultra Deep Field
(HUDF) main field dataset, as well as the same sky area as imaged by
ACS in the Great Observatories Origins Deep Survey (GOODS). In
particular, we used the optical images taken in the 4 ACS bands: F435W
(\bb), F606W (\vv), F775W (\ii), and F850LP (\zz).

We will briefly describe each dataset hereafter, but more details can
be found in \cite{giavalisco2004} and \cite{beckwith2006}.

\subsection{The Hubble Ultra Deep Field}

The main field HUDF dataset was obtained by combining all the
observations carried out during HST Cycle 12 (from September 2003 to
January 2004) under the two proposals 9978 and 10086 (P.I.:
S. Beckwith) with the aim of getting the deepest optical image of the
Universe using the Wide Field Camera (WFC) on HST/ACS.
The selected field is an 11 arcmin$^2$ sky area centered on
R.A. = $3^{\rm h}32^{\rm m}39^{\rm s}$, Dec. = $-27^{\circ}47'29''1$
(J2000.0) that resides within the Chandra Deep Field South (CDF-S).

The HUDF was imaged in four ACS optical bands: \bb, \vv, \ii, and
\zz.  The magnitude zero point, in AB magnitudes, the total exposure
time, and the depth ($5\sigma$ limiting magnitude within circular
apertures of 0\farcs35 diameter) for each filter are listed in Table
\ref{tab:param}.

\subsection{The Great Observatories Origins Deep Survey}
In 2003, version 1.0 (v1.0) of the GOODS images were released for both
the HDF-N and CDF-S fields acquired with HST/ACS as part of the GOODS
ACS Treasury program. In 2008, version 2.0 (v2.0) of the data was
released, which included the additional data acquired on the GOODS
fields during HST Cycles 12 and 13 looking for high-redshift Type Ia
supernovae (Program IDs 9727, 9728, 10339, 10340,
\citealt{riess2007}).  Each field is divided into sections,
i.e. images 8192 x 8192 pixels in size. A total of 17 and 18 sections
cover the HDF-N and CDF-S field respectively. These sections are
labeled according to their position in the field using a two digit
number.

The UDF GOODS dataset we used for the following analysis consists of
all the data located within the UDF pointing\footnote{The UDF GOODS
  v1.0 data release is available at
  ftp://archive.stsci.edu/pub/hlsp/udf/goods1/}.   Since a separate
  mosaic with GOODS v2.0 depth covering just the HUDF sky area did not
  exist, we combined the single tiles of the v2.0 data release to
create the image we needed.  In particular, we downloaded from the
MAST archive\footnote{The GOODS v2.0 dataset is available at
  http://archive.stsci.edu/pub/hlsp/goods/v2/} the images for Sections
13, 14, 23, 24, 33, and 34 in all the bands and combined them to get
the final images of the UDF sky area as seen by GOODS v2.0.  We made
available these images in the MAST archive as High-Level Science
Products\footnote{ The UDF GOODS v2.0 data release by V. Calvi is
  available ftp://archive.stsci.edu/pub/hlsp/udf/goods2/}.  For the
following analysis, we will consider only the v2.0 of the GOODS dataset
since it is slightly deeper than the previous version in the \vv, \ii,
and \zz-band.

As for the HUDF dataset, the magnitude zero point, the total exposure
time, and the depth for each filter are listed in Table
\ref{tab:param}.

\begin{table*}[ht]
\begin{center}
\caption{Properties of the datasets }
\label{tab:param}
\begin{tabular}{l|c|cc|cc}
\hline \hline
         &             & \multicolumn{2}{c|}{GOODS}    &  \multicolumn{2}{c}{HUDF}        \\
 Band    &  Zero Point & Exposure &  $5\sigma$ limiting & Exposure  & $5\sigma$ limiting   \\
         &  [AB mag]   & Time [s] &  magnitude [AB mag] & Time [s]  & magnitude [AB mag]    \\
\hline                  
\bb &    25.673 & 7200  & 28.30 & 134900 & 29.73\\
\vv &    26.486 & 5450  & 28.76 & 135300 & 30.14\\
\ii &    25.654 & 7028  & 27.72 & 347100 & 29.84\\
\zz &    24.862 & 18232 & 27.78 & 346600 & 29.18\\
\hline
\end{tabular}
\end{center}
NOTES: For each dataset and ACS filter the exposure time in seconds,
zero point in AB magnitudes,  and 5$\sigma$ limiting magnitude are
  listed. The depths were measured in circular apertures of
  $0\farcs35$ diameter.
\end{table*}

\section{Rescaling the rms maps}
When the images are characterized by variable noise, which is the case
for our data, for a proper error analysis SExtractor
\citep{bertinarnouts1996} needs weight maps as input, i.e. frames
having the same size as the science images that describe the noise
intensity at each pixel.  There are several options available (NONE,
BACKGROUND, MAP RMS, MAP VAR, MAP WEIGHT) and we choose to use the MAP
RMS  because it is not rescaled by the software, according to its
  manual\footnote{The SExtractor User's manual is available at
  www.astromatic.net/pubsvn/software/sextractor/trunk/doc/\\sextractor.pdf}.
We made use of the inverse-variance weight images ({\it *wht.fits})
produced by {\tt MultiDrizzle} to generate the rms maps.

According to its definition, the rms map is $\frac{1}{\sqrt{wht\;
    image}}$ but, since the errors on fluxes and magnitudes in
SExtractor depend directly on the weight image given as input, we
carefully checked the rms maps and derived a rescaling factor for each
of them.  This rescaling factor is needed to take into account the
correlated noise introduced during the drizzling process when a kernel
different from the ``point'' one is used \citep{casertano2000}.

We derived the rescaling factors for each dataset separately. We
selected 100 square regions with an area of 7 x 7 pixels with no
sources, i.e. just with background signal, from the \zz-band science
image of each dataset. Then, for all the bands, we computed the mean
signal characteristic of a $49\; {\rm pixels}^2$ area and the
associated rms error. Then, on the same 100 regions, we derived the
mean value in the corresponding rms map.  Finally, for each region, we
divided the root mean square of the signal in the science image by the
mean value of the signal in the rms map to get the rescaling factor
for the rms map.  The rescaled rms maps were then used as input weight
images when running SExtractor to get the catalogs of sources from the
different images.

As a further check we considered again the 100 background regions
selected before and computed the mean flux characteristic of a 7 x 7
pixel sky area and the rms associated with it. Then, we selected from
the catalogs all the sources with an ISO\_AREA in the range 47-51
pixels$^2$ and averaged the associated ISO\_FLUXERR values.  The two
values were supposed to be in good agreement, but we found they were
not. This means that SExtractor modifies, in some way, even the kind
of weight maps that are not supposed to be adjusted.  To overcome this
discrepancy and ensure that the SExtractor errors for each object are
not underestimated we iteratively changed the rescaling factor to get
a new rms map, re-ran SExtractor with this new rms map and, finally,
compared again the two values until we got a difference between them
that was less than 1\%.  The final rescaling factors for each filter
and dataset are recorded in Table
\ref{tab:rescaling_factors_final}. Note that a rescaling is needed
also for the HUDF \ii\/ and \zz-band images that were obtained with the
point kernel and would be expected to be safe from correlated
  noise components. This was claimed already by \cite{oesch2007}.

\begin{table*}[]
\begin{center}
\caption{Final rescaling factors}
\label{tab:rescaling_factors_final}
\begin{tabular}{l|cccc}
\hline\hline
Dataset     & \bb & \vv & \ii &\zz\\ 
\hline
GOODS v2.0 &  2.055 & 3.150 & 2.900 & 3.345\\
HUDF       &  1.220 & 2.220 & 2.280 & 1.670\\
\hline
\end{tabular}
\end{center}
NOTES: The first estimate for the rescaling factors was derived
comparing the typical rms associated to the background signal in the
science images and the mean signal derived from the rms map over 100
selected 7 x 7 pixel regions. Then, we derived the final values
following an iterative approach based on the comparison between the
typical rms associated to a background region and the SExtractor
ISOFLUX\_ERR derived for objects with the same total area.
\end{table*}

\section{Sample Selection}
\label{sec:selection}

To study the possible effect of cosmological dimming, we need to
compare the magnitude of the same high-redshift sources derived from
surveys that imaged the same sky area, but are characterized by
different depth.  Among our choice of data, the GOODS dataset is the
shallowest and, essentially, determines the kind of sources we could
study, as well as their number, since we are interested in objects
detected in both the surveys. 

High-$z$ galaxies in deep surveys are usually selected on the basis of
the Lyman Break technique.  Dropout galaxies show little or no flux at
wavelengths shortward of the rest-frame Ly$\alpha$ line due to strong
absorption by intergalactic Hydrogen.  In detail, there is no flux
below 912 \AA\/ rest-frame due to continuum absorption. At $z \gtrsim
3$, as $z$ increases the Ly$\alpha$ forest becomes increasingly opaque
so that the effective break moves to 1216 \AA\/ rest-frame. 
At $z\sim4$, for example, this results in a lack of flux in the
\bb-band, i.e. no detection, associated with a clear detection in the
redder bands, in our case \vv, \ii, and \zz.  A comprehensive study of
Lyman Break galaxies at $z\sim4-5$, including the determination of the
fraction of interlopers entering the sample, was conducted by
\cite{huang2013} who studied the bivariate size-luminosity relation
using HUDF and GOODS data.

We ended up considering only the \bb-dropout population, i.e.
candidate high-$z$ galaxies at $z\sim4$, and we discarded the analysis
on \vv$\,$ and \ii-dropouts because the two samples included only two
and one candidate, respectively, and were, indeed, not statistically
significant.  We selected our sources by running SExtractor version
2.8.6 \citep{bertinarnouts1996} in dual image mode within each dataset
using the \zz-band image for the detection and performing the
photometry on all the four bands.  We used the same SExtractor
parameters stated in \cite{beckwith2006} since they were optimized for
the pixel scale and Point Spread Function (PSF) typical of the
datasets.  The threshold for detection and analysis (DETECT\_THRESH
and ANALYSIS\_THRESH) was set to be 0.61 within an area of at least 9
contiguous pixels (DETECT\_MINAREA).  We selected a total of 32
deblending subthresholds (DEBLEND\_NTHRESH), a contrast parameter of
0.03 (DEBLEND\_MINCONT), and a Full Width Half Maximum (SEEING\_FWHM)
of $0''.09$.  Moreover, we used the {\it goods.conv} filter
(FILTER\_NAME) optimized for GOODS data\footnote{ The filter named {\it
  goods.conv} is a 9 x 9 pixel  convolution mask of a Gaussian PSF
with FWHM = 5.0 pixels.}.  We set the WEIGHT\_TYPE to MAP\_RMS and
WEIGHT\_IMAGE to be the final rms map derived applying the rescaling
factors listed in Table \ref{tab:rescaling_factors_final}. Since the
detection is affected by the rescaling factor as well, within each
dataset we rescaled the 0.61 value we used as DETECT\_THRESH and
ANALYSIS\_THRESH according to the rescaling factor in the \zz-band.
We applied the following color selection criteria, as stated in
  \cite{beckwith2006}:
\begin{equation}
\begin{aligned}
 B_{435}-V_{606} &> (1.1 + V_{606}-z_{850} )\\
 B_{435}- V_{606} &> 1.1\\
V_{606} -z_{850} &> 1.6\\
(S/N)_{V_{606}} &> 5\\
(S/N)_{i_{775}} &> 3
\end{aligned}
\end{equation}

\begin{figure}[]
\centering
\includegraphics[scale=0.43]{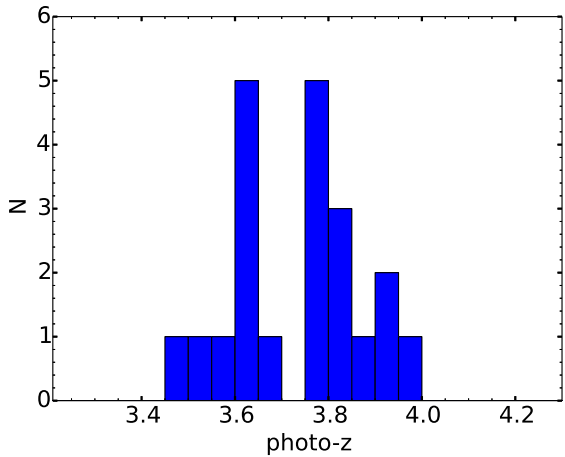}
\caption{Distribution of the photometric redshift values obtained
  running the BPZ code \citep{benitez2000}. The narrow distribution
  matches the redshift window associated with \bb-dropout galaxies
  ($3.4 \lesssim z \lesssim 4.4$). }
\label{fig:photo-z}
\end{figure}

\noindent  The colors are calculated from the MAG\_ISO in each filter and
the detection $S/N$ is derived from the FLUX\_ISO and FLUXERR\_ISO
parameters according to \cite{stiavellibook}
\begin{equation}
\frac{S}{N}=\frac{\rm FLUX\_ISO}{\rm FLUXERR\_ISO}
\end{equation}
These criteria permit a selection of candidates that lie in the
redshift range $3.4 \lesssim z \lesssim 4.4$ and, according to
\cite{huang2013}, the fraction of lower redshift interlopers is below
8\%.  Following \cite{vanzella2005}, we checked the GOODS/CDF-S
spectroscopic
catalog\footnote{http://archive.eso.org/wdb/wdb/vo/goods\_CDFS\_master/form}
looking for tabulated spectroscopic redshifts. Out of our sample, only
3 objects had a spectroscopic follow-up, but all of them are confirmed
\bb-dropouts since they have $z_{spec}$ equal to 3.887 (C), 3.797 (A),
and 3.604 (A), respectively.\footnote{For ESO-GOODS spectroscopy (A)
  means secure redshift determination, (B) a likely redshift
  determination, and (C) a tentative redshift determination.} We also
used the HUDF data to derive the photometric redshifts by running the
Bayesian Photometric Redshifts (BPZ) code by
\cite{benitez2000}. Despite the use of only four photometric bands, we
obtained a good fit for each object in the sample, with a probability
always higher than 95\%. As shown in Figure \ref{fig:photo-z}, the
distribution of photometric redshifts for the sample matches the
redshift window associated with \bb-dropout galaxies.

 The final samples contain 74 and 428 dropout galaxies for the GOODS
 v2.0 and HUDF dataset, respectively. It should be noted that the area
 covered by the GOODS images of the HUDF sky area is almost twice the size of
  the HUDF sky area, thus several \bb-dropouts lie in regions of
 the image where there is not coverage in the HUDF images.  In
   detail, out of the 74 \bb-dropouts detected in GOODS, 48 lie
   outside the actual HUDF footprint. After obtaining the catalogs of
 \bb-dropouts matching in both datasets, we visually inspected all the
 sources and rejected any spurious detections.  A total of five
 sources were discarded.

\section{Results}

\begin{figure*}[]
\centering
\includegraphics[scale=0.46]{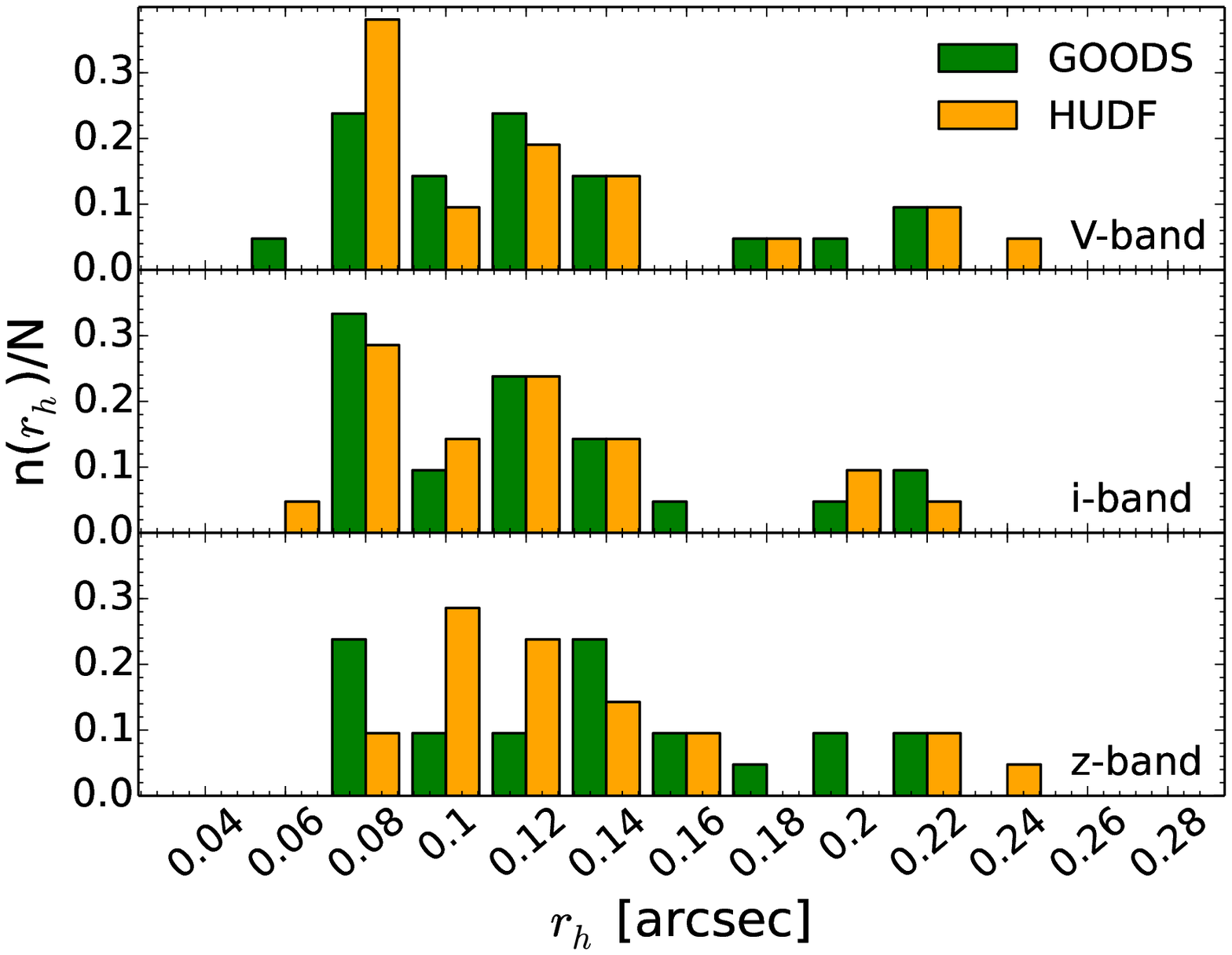}
\includegraphics[scale=0.46]{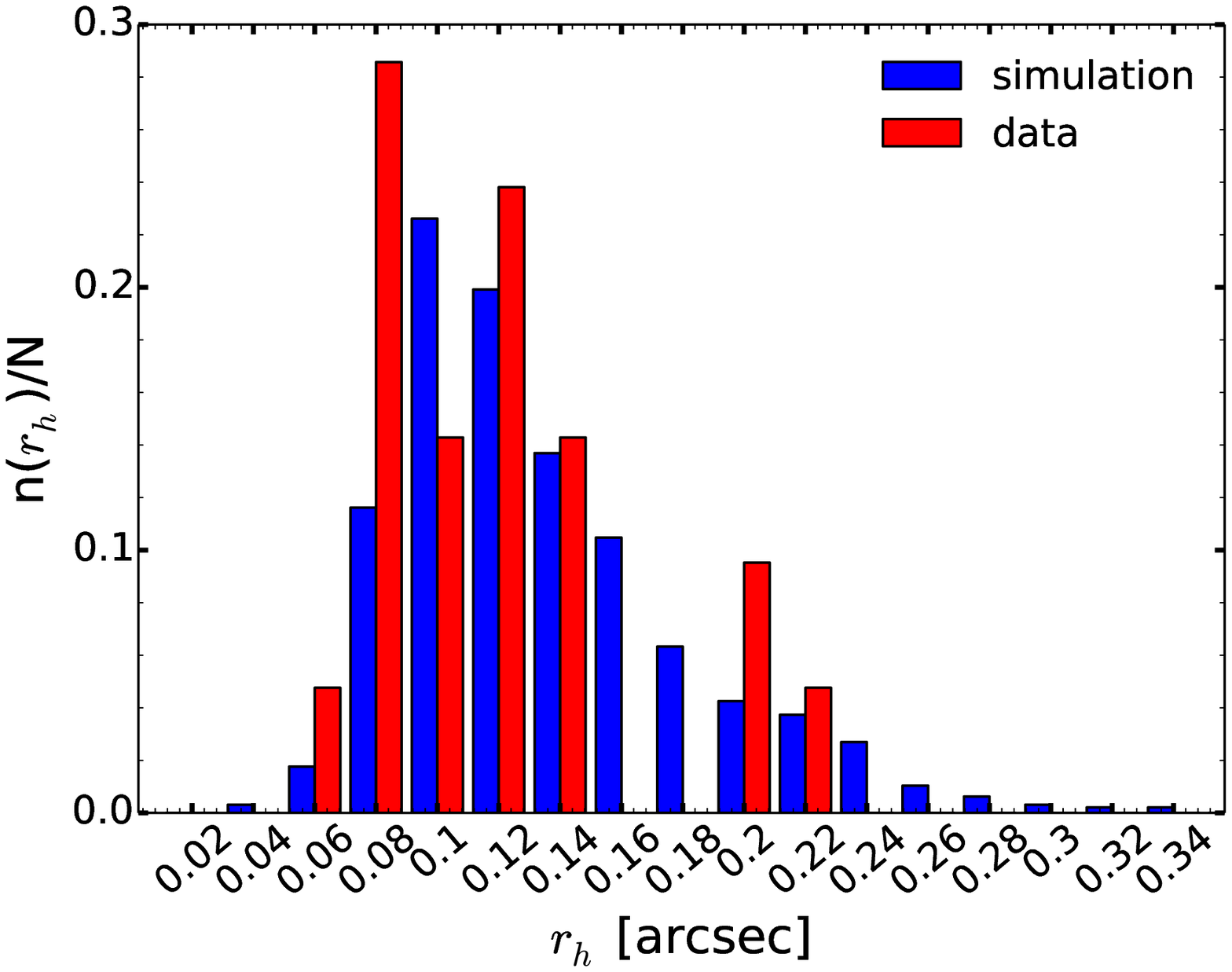}
\caption{Left panels: for each detection band the bar plots
  compare the normalized distribution of values for the half-light
  radius $r_h$ obtained running SExtractor on GOODS-depth (in green)
  and HUDF-depth (in orange) images. Right panel: bar plots comparing
  the normalized distribution of half-light radii values from \ii-band
  data (in red) and the simulations (in blue). In detail, we plotted
  an example of the distribution of half-light radius values recovered
  by running SExtractor on mock images. All plots are characterized by
  a bin size of 0.02 arcsec and n($r_h$) and N indicate the number of
  galaxies per bin and the total number of objects analyzed (i.e. 21
  in our sample and 1000 in each simulation run), respectively. }
\label{fig:radii}
\end{figure*}

Comparing the HUDF data with GOODS v2.0, we obtained a sample of 21
matching objects within the HUDF footprint that were selected as
\bb-dropout candidates independently in the two datasets.   The
  half-light radii $r_h$, derived running SExtractor, are plotted in
  the left panels of Figure \ref{fig:radii}. The radii derived from
  GOODS-depth images and those coming from the HUDF-depth ones are in
  good agreement.  

\begin{figure*}[!h]
\centering
\includegraphics[scale=0.42]{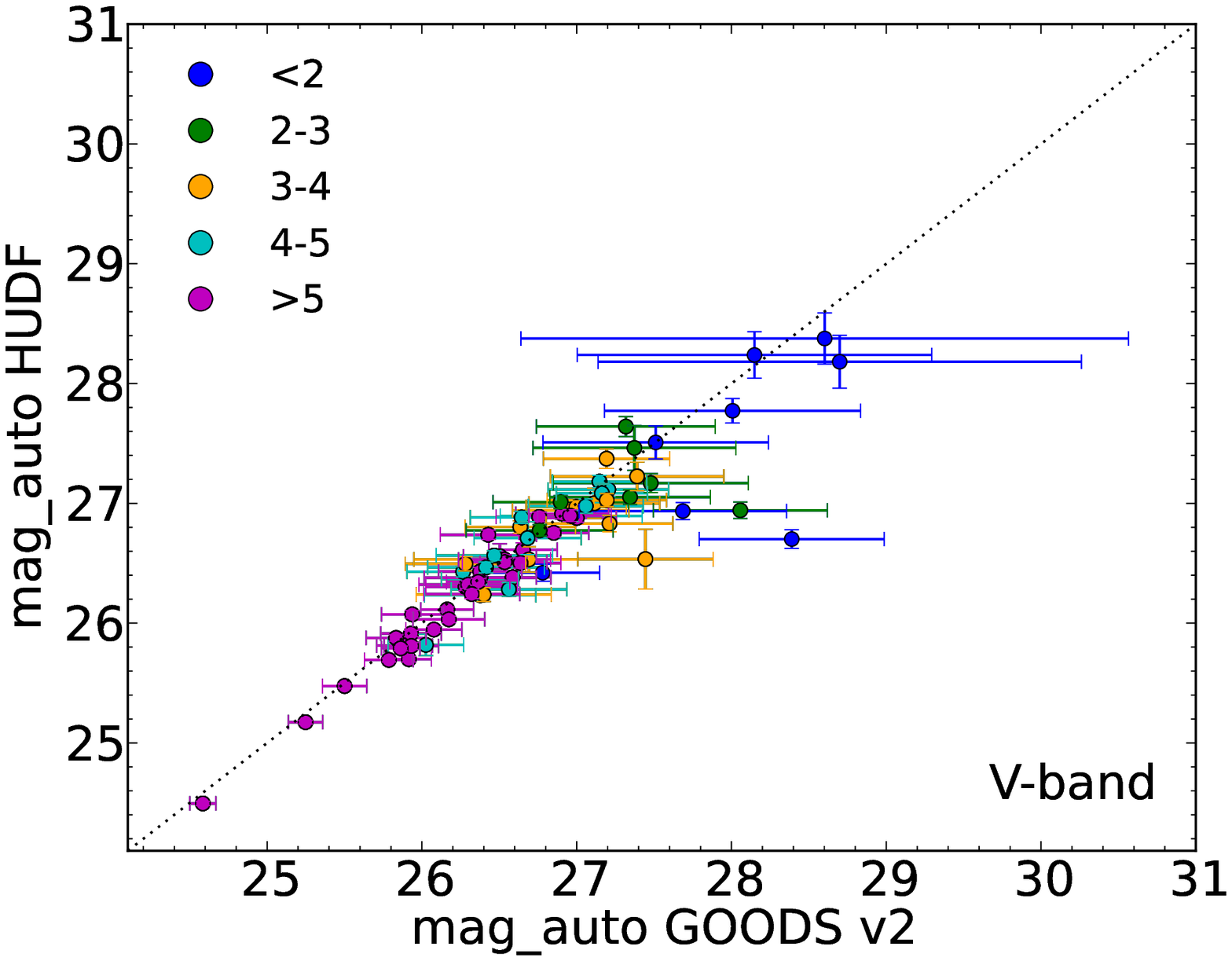}
\includegraphics[scale=0.42]{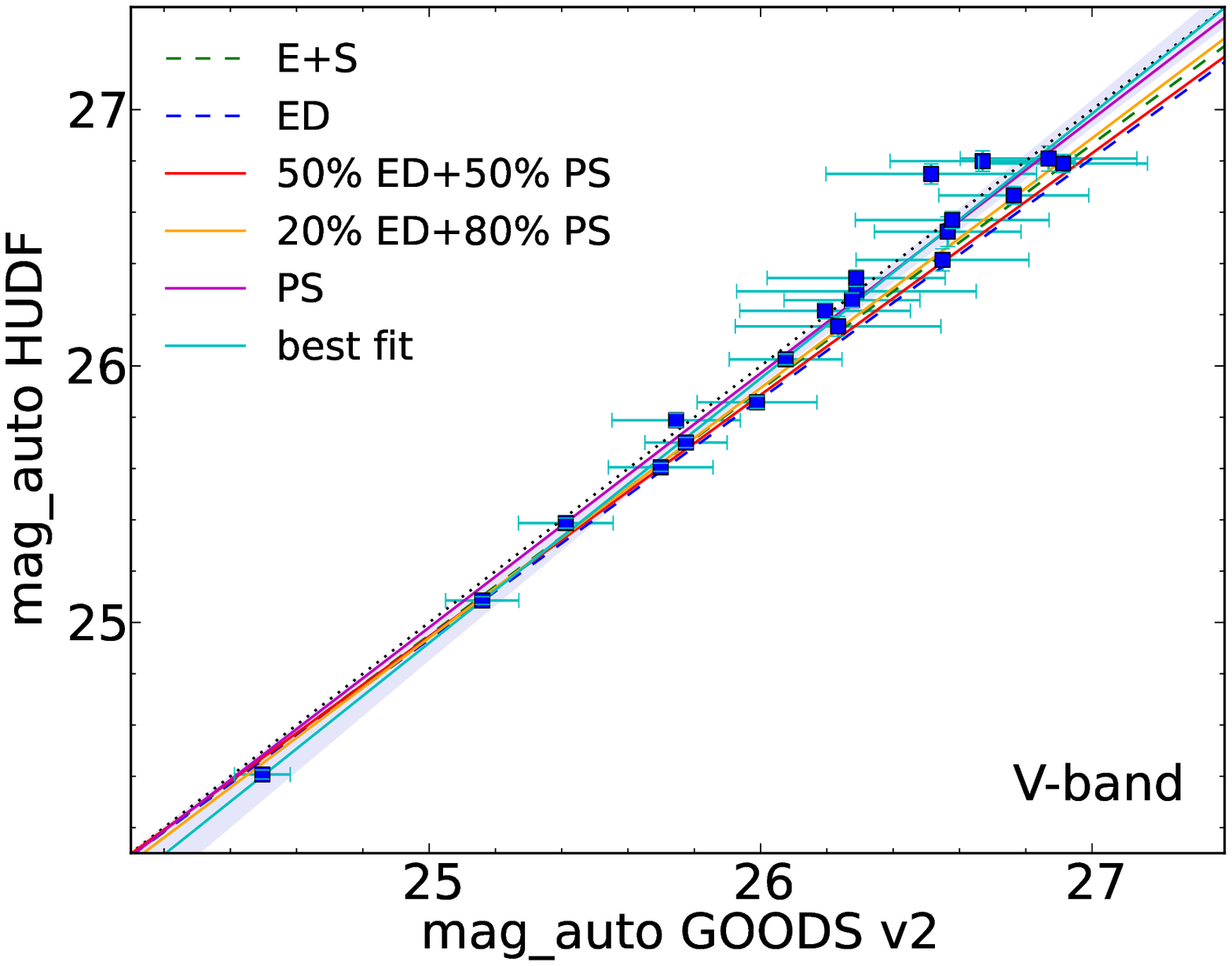}
\includegraphics[scale=0.42]{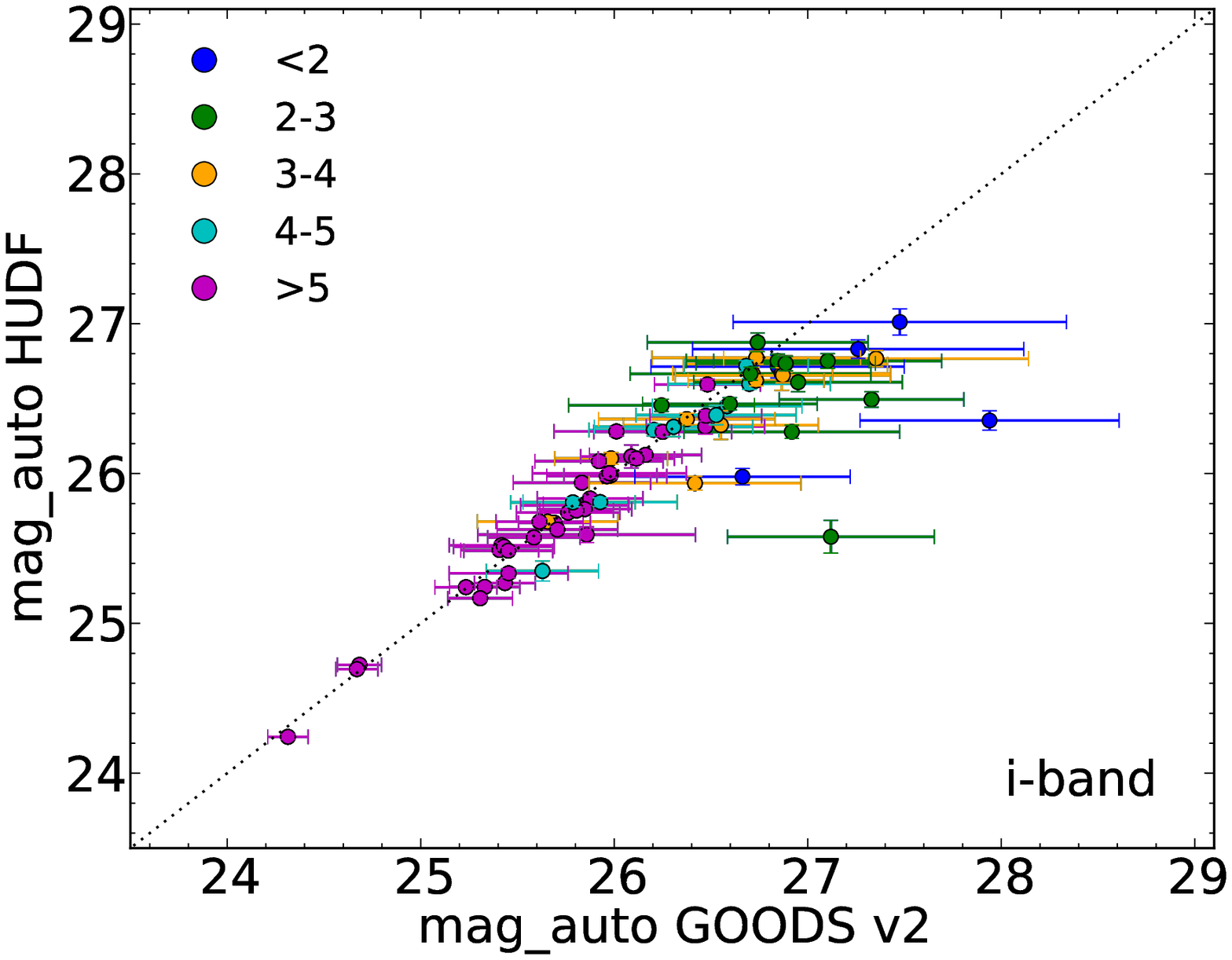}
\includegraphics[scale=0.42]{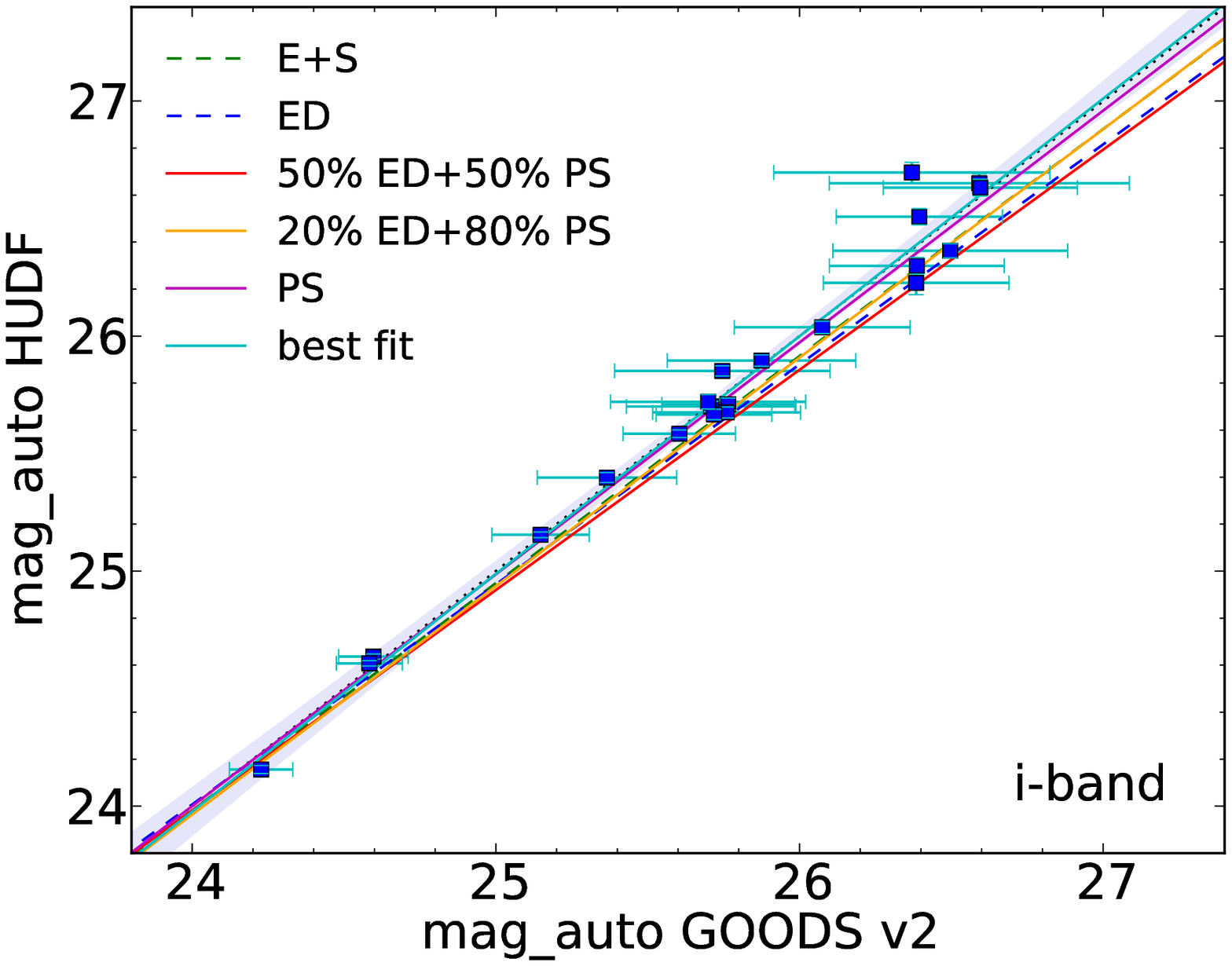}
\includegraphics[scale=0.42]{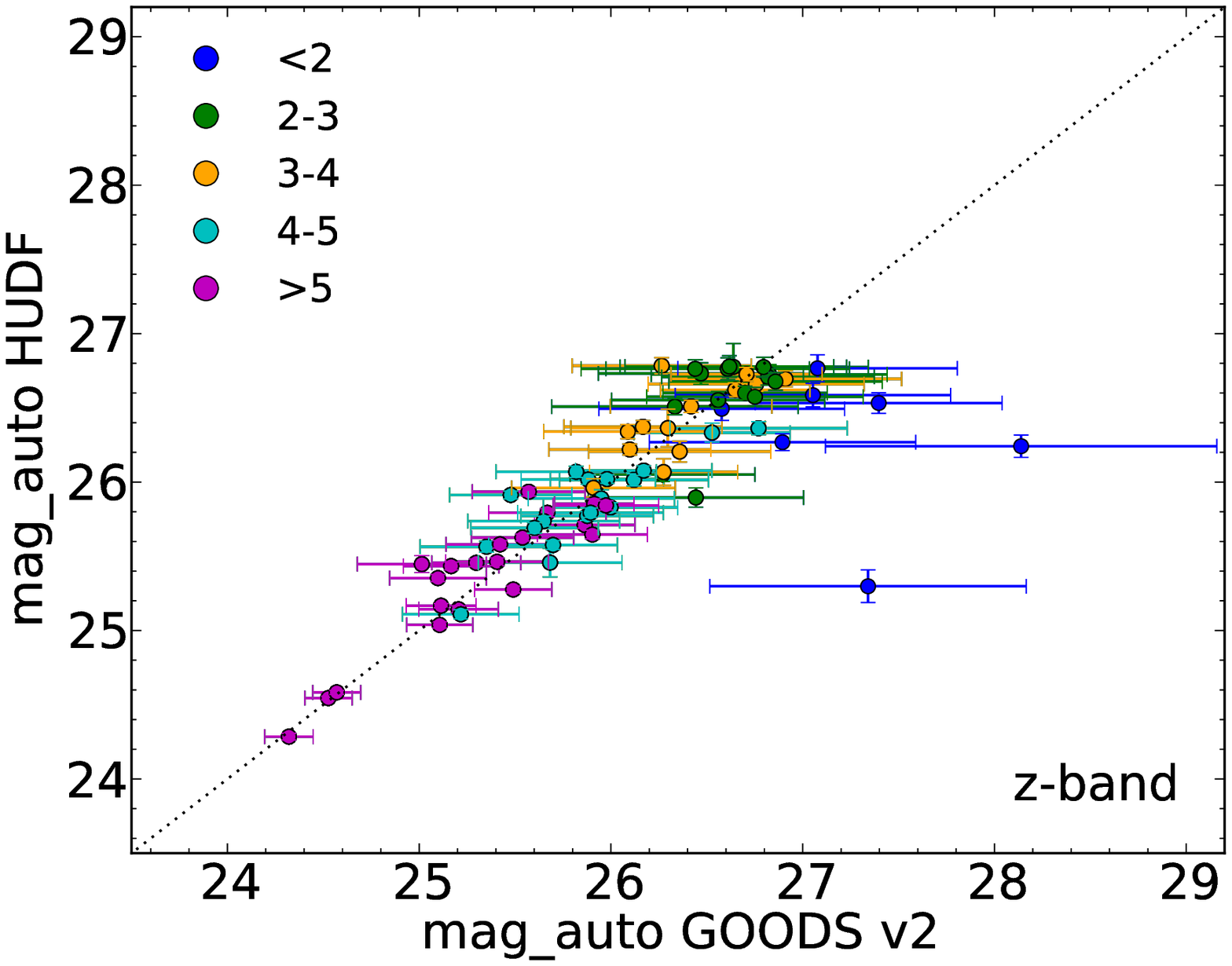}
\includegraphics[scale=0.42]{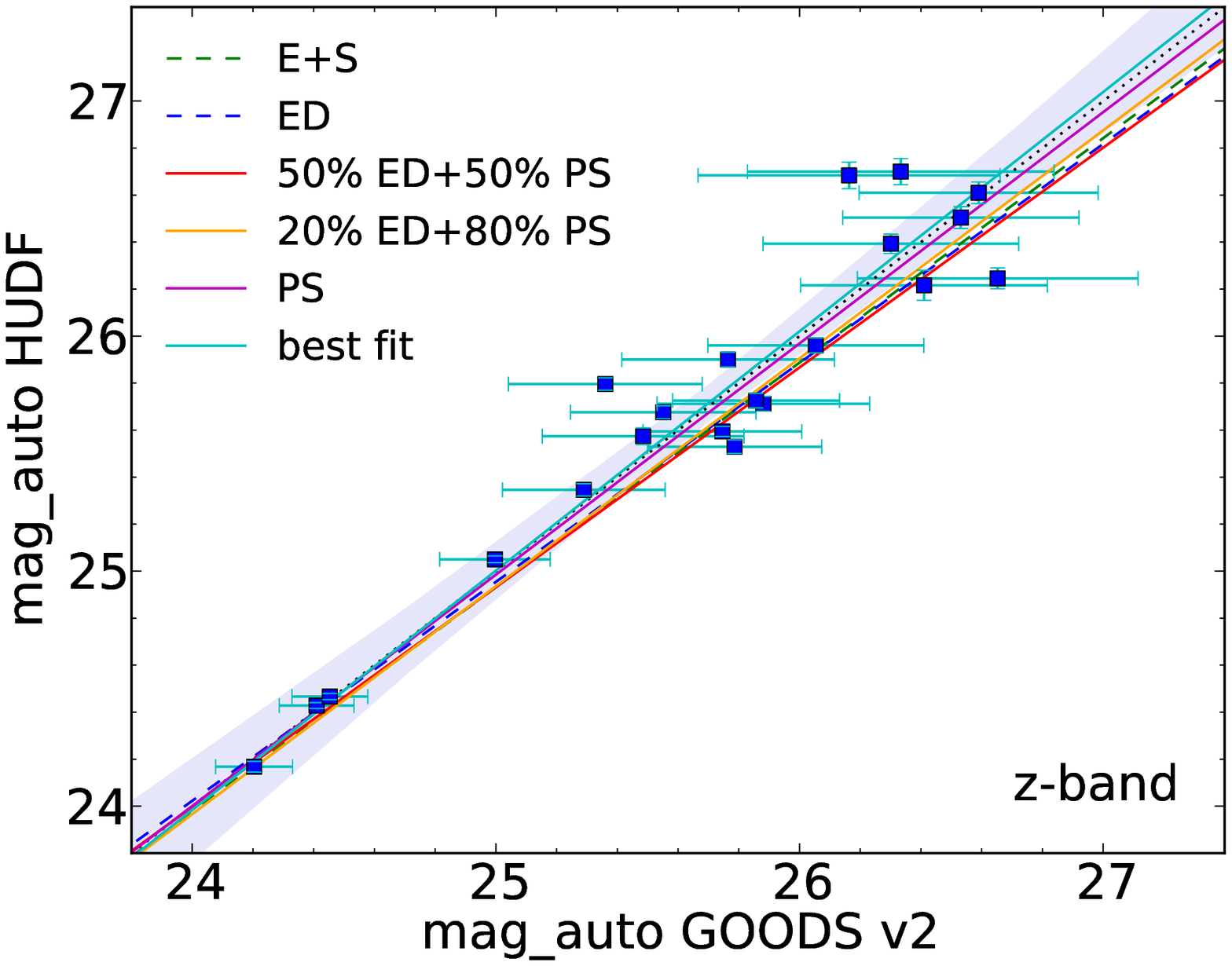}
\caption{ Total magnitude derived from SExtractor MAG\_AUTO parameter
  in the \vv (top panels), \ii(central panels), and \zz-band (bottom
  panels) from the GOODS v2.0 dataset compared to the one derived from
  the HUDF. Left panels refer to all the objects selected in HUDF as
  \bb-dropouts that have a detection in GOODS v2.0. As tabulated in
  the legend, different colors refer to different cuts in $S/N$ in
  GOODS if no color selection is applied in GOODS. Right panels show
  the sample, consisting of 21 \bb-dropouts, satisfying all the
  selection criteria in both HUDF and GOODS v2.0.  Since the error
    bars on the y-axis are small they are often hidden by the symbols.
  The colored lines indicate the trend if cosmological dimming is
  acting depending on the way the flux is distributed among the
  galaxy, while the dotted black line shows the 1:1 trend. The shaded
  area shows the 95\% confidence band around the best fit (solid cyan
  line). In the legend E, S, ED, and PS indicate ellipticals,
    spiral galaxies, exponential disk, and point source,
    respectively.}
\label{plot_GOODS_HUDF_viz}
\end{figure*}

For all the \bb-dropouts in our sample, we plotted the total
magnitudes and errors, given by SExtractor MAG\_AUTO and MAGERR\_AUTO
parameters, respectively, and derived the best linear least-squares
fit $y=A+B\cdot x$ considering the error bars in both coordinates
using the {\tt IDL\footnote{Interactive Data Language is distributed
    by Excelis Visual Information Solutions.}} {\tt fitexy.pro}
routine (right panels of Figure \ref{plot_GOODS_HUDF_viz}). The best
fit parameters are tabulated in Table \ref{tab:fit}.

As a further check, we investigated the behavior of all the
\bb-dropouts selected in HUDF, with no regard for their
identification in GOODS v2.0.  In particular, we selected all the
\bb-dropouts detected in HUDF with a total magnitude in the \zz-band
brighter than \zz=26.8 mag, the value that defines the completeness of
the GOODS dataset (50\% complete). 79 \bb-dropouts satisfy this
criterion and we focused on them for the following analysis.
Discarding the cuts in $S/N$ for the GOODS data, there are 75 objects
selected as \bb-dropouts in HUDF that have a detection in GOODS v2.0,
two of them were discarded as each of them is clearly detected as two
distinct sources in GOODS. Out of the 73 remaining objects we ran a
separate analysis considering four different cuts in $S/N$  in
  GOODS-depth images only: $S/N>2$, $S/N>3$, $S/N>4$, $S/N>5$. For
each selection we derived the best fit, as done for the sample of
matching \bb-dropouts.

The left panels of Figure \ref{plot_GOODS_HUDF_viz} summarize our
results plotting the total magnitudes in GOODS and HUDF on the basis
of the $S/N$ cut  for the different bands.  As can be seen from
Table \ref{tab:fit}, there is no significant change in the trend of the
linear fit, even including objects with lower $S/N$ detection, 
  showing that deeper images do not reveal any low-SB features, hard to
  detect due to cosmological dimming, in the sample galaxies.  When
including the objects with very poor or no detection in at least one
of the bands used to identify objects, there could be an effect due to
SB dimming, but we are not able to disentangle this effect from the
one simply due to the detection limit of GOODS.

\begin{table*}[!th]
\begin{center}
\caption{Best linear least-square fit}
\label{tab:fit}
\begin{tabular}{l|ccc|ccc|ccc}
\hline 
\hline
Selection &  \multicolumn{3}{c|}{\vv} & \multicolumn{3}{c|}{\ii}    &  \multicolumn{3}{c}{\zz}  \\
Criteria  & \# obj   & A & B   & \# obj & A & B  & \# obj  & A & B \\
\hline
$S/N>5$ + color & 21 & -0.88 & 1.03 & 21 &-0.28 & 1.01 & 21& -0.45 & 1.02\\
$S/N>5$         & 32 & -1.04 & 1.04 & 36 &-0.59 & 1.02 & 20& -1.13 & 1.04\\
$S/N>4$         & 46 & -1.13 & 1.04 & 45 &-0.46 & 1.02 & 38& -0.48 & 1.02\\
$S/N>3$         & 58 & -0.93 & 1.03 & 57 &-0.17 & 1.01 & 51& -0.67 & 1.03\\
$S/N>2$         & 65 & -0.86 & 1.03 & 68 & 0.30 & 0.99 & 66& -0.51 & 1.02\\
All             & 73 & -0.58 & 1.02 & 73 & 0.56 & 0.98 & 73& -0.03 & 1.00\\
\hline
\end{tabular}
\end{center}
NOTES: Parameters for the best linear least-square fit in the form
$y=A+B \cdot x$.  In the fit equation $x$ and $y$ are MAG\_AUTO
  values in GOODS-depth and HUDF-depth, respectively. A and B were
derived comparing the total magnitudes in the \vv, \ii, and \zz-band
for the sample of matching sources taking into account the error bars
associated to the magnitude values.  Different rows refer to different
selection criteria and, consequently, the number matching of objects
varies.
\end{table*}

\section{Simulations}

Our results suggest no effect due to cosmological dimming when
comparing deep sky surveys with different depth. To complete our study,
we wanted to show that the difference in depth between our two
datasets would be enough to detect Tolman dimming effects if galaxies
were characterized by a smooth light distribution at the resolution of
HST. To this aim, we ran Monte Carlo simulations reproducing the
\bb-dropout sample in order to derive the expected trend in the
magnitude-magnitude plot in the case of cosmological dimming affecting
the SB of the sources.

 In detail, we created a 4000 x 4000 pixel image in each band into
 which we inserted 1000 mock galaxies with apparent magnitudes ranging
 from 24.5 to 28.5 mag, axis ratio $b/a$ and effective radius ${\rm
   r_e}$ distributed according to the findings of \cite{ferguson2004},
 and position angle and x,y positions randomly drawn from a flat
 distribution.  We used the {\tt noao.artdata.mkobject} package in
 {\tt IRAF}\footnote{IRAF is distributed by the National Optical
   Astronomy Observatories, which are operated by the Association of
   Universities for Research in Astronomy, Inc., under cooperative
   agreement with the National Science Foundation.} to make the
 artificial galaxies. The right panel of Figure \ref{fig:radii} shows
 a comparison between the normalized distribution of the half-light
 radius values from one of our simulations and our \bb-dropout
 sample. Each value of $r_h$ used to create a mock galaxy was drawn
 from the size distribution presented in \cite{ferguson2004}, that is
 based on measurements performed by SExtractor on \ii-band images.
 The distribution of the half-light radii, recovered running
 SExtractor on the i-band image of our dataset, is in good agreement
 with those recovered from the simulated galaxies.

In the first set of simulations we reproduced the images in each band
assuming that all the galaxies have just one component.  First we
created a mock image containing only disk-like galaxies,
i.e. late-type, with a surface brightness exponential profile.  We
convolved the mock frame with the corresponding PSF model obtained
running {\tt Tiny Tim}\footnote{The Tiny Tim web interface is
  available at http://www.stsci.edu/hst/observatory/focus/TinyTim.}
\citep{krist2011}.  This tool models the PSF specific to a chosen
channel for any of the HST imagers and bands.  We obtained three
different PSF models, one for each detection band.  Finally, we added
a random noise frame with the same rms as the one measured on the
science image in the corresponding band.

We also created a set of mock images assuming that 25\% of the galaxies
have a De Vaucouleurs profile (i.e. $r^{1/4}$), and the remaining an
exponential one. This is broadly compatible with \cite{ferguson2004}
who found that, among the \bb-dropout sample, 78\% of the galaxies are
classified as late-type. As previously done, we convolved all mock
objects with the PSF by {\tt Tiny Tim} and added random noise.

In both cases, i.e. exponential and mixed profiles, the
effect of surface brightness dimming is expected to be large
since the objects are extended and  the flux is mostly distributed in
extended structures.

In the third set of simulations, we used mock galaxies that have a
double component. Specifically, we used an exponential disk and a
central point source. We used the same PSF image we got from {\tt Tiny
  Tim} to convolve the frames as the model for the point source and we
convolved only the exponential disks with the typical ACS PSF. As done
for the simulations previously described, we added the random
noise reproducing the background of the science image in the selected
band.  In detail, we simulated galaxies with 80\% of the flux
enclosed in the central point source and 20\% in the extended
exponential disk or with the flux equally split among the two
components.

Finally, we ran the last set of simulations modeling each mock galaxy
with the PSF by {\tt Tiny Tim}, i.e. we created point sources that are
not expected to be affected by the curvature $(1+z)^2$ component of
cosmological dimming. In this case we did not convolve the synthetic
frame with the PSF before adding the background noise.

For each set of simulations we performed 50 iterations for each band.
 Then, we ran SExtractor on each mock image with the same parameters
used for the detection in the real data. Finally, we compared the
GOODS-depth and HUDF-depth magnitudes for the mock galaxies from the
output catalogs.  For each realization of the simulations we derived a
linear fit describing the expected trend the GOODS-depth magnitude vs
HUDF-depth magnitude plot, then we averaged the A and B values over
all the 50 realizations in order to derive the best fit parameters
listed in Table \ref{tab:profiles_best_fit}.

To determine whether or not the trends derived from the simulations
are in agreement with our data, we computed the normalized residuals
for each light distribution profile.  In detail, we derived the
distance between each data point and the linear fit obtained from a
set of simulations and divided it by the corresponding error.  For
compact sources, for which no significant dimming is expected, we
predict the distribution of the residuals to resemble the one obtained
with respect to the 1:1 line.  Observed objects tend to lie
systematically to the left of the model lines (right panels of Figure
\ref{plot_GOODS_HUDF_viz}). The residual distribution captures this
systematic effect which, instead, can be missed by tests relying only
on the squares of the residuals.  We performed a Kolmogorov-Smirnov
non-parametric statistical test comparing the distribution of the
residuals derived from each set of simulations and the one from the
data.  The values of the maximum vertical deviation between the two
cumulative curves, namely the $D$ parameter, and the corresponding
probability $p$\footnote{In the Kolmogorov-Smirnov test $p$ is the
  probability of two random samples having a $D$ parameter as large or
  larger than the measured one.}  are listed in Table
\ref{tab:KS_statistic}.  If the $p$ value is small, the two groups
were sampled from populations with different distributions.  We can
reject the null hypothesis (i.e. the two distributions are the same)
with a 95\% confidence for all the realizations characterized by
$p<0.05$.  As can be seen the expected trend in the cases of flux
distributed in extended sources ($r^{1/4}$ and exponential profile)
can be easily ruled out in the \vv\; and \ii-band, while the trend for
point sources is in agreement with our findings, suggesting that the
light from high-$z$ sources mostly comes from compact objects rather
than extended ones.  It should be noted that diffuse models are ruled
out with lower significance in the \zz-band and this could hint at a
diffuse component starting to be more significant in the redder bands.
A compact structure at high-$z$, at least in the bluer bands with a
detection, is in agreement with the results obtained by
\cite{williams2014} for $z\sim3 $ LBGs.

As a further result, the simulations involving point sources permitted
us to verify the findings of \cite{ashby2013} who claimed that the
total magnitude recovered by SExtractor is, on average, 0.05 mag
fainter than the real value.  We found all the best fit lines
  derived for the point source simulations to be slightly shifted in
  the x direction and that, introducing the  0.05 mag correction
  stated by \cite{ashby2013}, the fit was closer to the expected trend
  determined by the 1:1 line.

\begin{table*}[]
\begin{center}
\caption{Linear least-square fit from Monte Carlo simulations}
\label{tab:profiles_best_fit}
\begin{tabular}{l|cc|cc|cc|cc|cc}
\hline
\hline
& \multicolumn{2}{c|}{Mix (E+S)} & \multicolumn{2}{c|}{exp profile} & \multicolumn{2}{c|}{point source}  & \multicolumn{2}{c|}{20\% exp profile} & \multicolumn{2}{c}{50\% exp profile} \\
Band &  A & B & A & B & A & B  & A & B & A & B\\
\hline
\vv &  0.70 & 0.97 & 1.48 & 0.94 & 0.20 & 0.99   & 0.59 & 0.97 & 1.39 & 0.94 \\ 
\ii &  0.74 & 0.97 & 1.54 & 0.94 & 0.33 & 0.99   & 0.64 & 0.97 & 1.49 & 0.94 \\
\zz &  1.01 & 0.96 & 1.67 & 0.93 & 0.39 & 0.98   & 0.69 & 0.97 & 1.75 & 0.93 \\
\hline
\end{tabular}
\end{center}
NOTES: Best linear least-square fit parameters derived for each of the
three detection bands averaging over 50 realizations of Monte Carlo
simulations. We considered different scenarios on the basis of the
distribution of the total flux: a mix of galaxies (25\% with
  $r^{1/4}$ profile and 75\% described by an exponential disk), an
  exponential disk, a point source modeled using the PSF image created
  using {\tt Tiny Tim}, and, finally, multi-component objects with the
  flux divided into an exponential disk and a point source modeled
  using {\tt Tiny Tim} PSF. In particular we considered two cases:
  either the light equally split among the disk and point source
  components or 20\% of the light in the exponential disk and 80\% in
  the central point source.
\end{table*}

\begin{table*}[]
\begin{center}
\caption{Results from the Kolmogorov-Smirnov statistic}
\label{tab:KS_statistic}
\begin{tabular}{l|cc|cc|cc}
\hline
\hline
& \multicolumn{2}{c|}{\vv-band} & \multicolumn{2}{c|}{\ii-band} & \multicolumn{2}{c}{\zz-band}\\
Profile  & $D$ & $p$ & $D$ & $p$ & $D$ & $p$\\
\hline
Mix (E+S)             & 0.57     &  1.08e-03 &  0.43 &  0.03     & 0.42 & 0.03  \\
Exponential disk      & 0.67     &  7.13e-05 &  0.57 &  1.08e-03 & 0.33 & 0.15  \\    
50\% Disk + 50\% PS   & 0.62     &  2.90e-04 &  0.66 &  7.10e-05 & 0.43 & 0.03 \\ 
20\% Disk + 80\% PS   & 0.52     &  3.59e-03 &  0.48 &  0.01     & 0.38 & 0.07  \\   
Point source          & 0.24     &  0.53     &  0.29 &  0.30     & 0.14 & 0.97  \\
\hline
\end{tabular}
\end{center}
NOTES: Results from the Kolmogorov-Smirnov test obtained comparing the
distribution of the residuals derived from each set of simulations and
the one from the data. The values of the $D$ parameter and the
corresponding probability $p$ are listed for all the models and bands,
as labeled.  When $p<0.05$ the distribution of the residuals differ
from the expected one with a 95\% confidence, so the model can be
ruled out as a possible fit for our data.
\end{table*}

\section{Conclusions}

To study the effects of surface brightness dimming on high-redshift
galaxies, we compared catalogs of candidate \bb-dropouts obtained
running SExtractor on  deep images taken with ACS as part of the 
GOODS and HUDF projects.
In particular, considering the total magnitude, given by the MAG\_AUTO
parameter, we found no systematic differences between GOODS and the
deeper images from HUDF. In detail, our sample of
\bb-dropouts suggests that there is no trend in the magnitude of
galaxies becoming brighter in the HUDF, as already stated by
\cite{beckwith2006}.

If most of the light of faint dropout galaxies was in compact knots,
one would expect little or no SB dimming. In contrast, if a substantial
fraction of the light is in a diffuse component, such components would
be detected in part in the HUDF and missed in GOODS. Our direct
comparison shows that galaxies detected in GOODS do not become
significantly brighter in deeper images.  This finding
suggests that most of their light is compact, as claimed by
\cite{bouwens2004size} who found that the principal effect of depth in
a galaxy survey is to add galaxies at fainter magnitudes, and not to
significantly add to the sample galaxies with bigger sizes. Furthermore,
our findings are in agreement with those by \cite{williams2014}.
Studying the morphology of these galaxies in the rest-frame UV and in
the optical, they found a very compact main structure and basically no
diffuse component, even in deep stacks.

Comparing the data to simulations reproducing the effect of
cosmological dimming for different SB profiles, we found that our
trends can not be ascribed to extended structures, such as $r^{1/4}$
profiles or exponential disks. However, diffuse models are ruled out
with lower significance in the reddest band, i.e. \zz-band, which could
hint at a diffuse component starting to be more significant.

The study of cosmological SB dimming is also important since it could
affect our prediction of what JWST can observe at higher redshifts,
where younger galaxies may exhibit a larger fraction of
clumpiness. Our direct comparison shows that galaxies detected in
GOODS do not become significantly brighter in the HUDF. This suggests
that most of their light is compact and hints at the fact that JWST
will likely not find extensive diffuse star forming components.  This
is also promising as it suggests that JWST will not be confusion
limited at least in the shortest bands.  

Whether or not galaxies have a red diffuse component remains an open
question and will require further investigations.
 
\acknowledgments{\noindent We acknowledge useful conversations with
  Harry Ferguson, Mike Disney, Laura Watkins, Michele Cignoni, and
  Gianluca Castignani. We thank the anonymous referee for his/her
  insightful comments that improved the paper. This work has been
  supported by JWST IDS grant J1597. VC is supported by a grant
  ``Borsa di studio per l'estero, bando 2013'' awarded by ``Fondazione
  Ing. Aldo Gini'' in Padua (Italy). AP acknowledges the financial
  support of grants 0A02-4807/12, 60A02-5857/13, CPDA133894 of Padua
  University.}

\noindent Facilities: \facility{HST(ACS)}.

\bibliographystyle{apj}

\bibliography{valebib} 

\end{document}